\documentclass{optica-article}

\journal{opticajournal} 

\articletype{Research Article}

\usepackage{lineno}
\usepackage{gensymb}

\usepackage{xcolor}
\usepackage{soul}

\begin{document}

\title{Rubidium-Doped KTiOPO$_{4}$ Waveguides as a Dual-Type Photon Pair Source}

\author{Patrick Hendra,\authormark{1,2,*} Josué R. León-Torres,\authormark{1,2,3} Valerio Flavio Gili,\authormark{1} and Markus Gräfe\authormark{1,3}}

\address{\authormark{1}Fraunhofer Institute of Applied Optics and Precision Engineering IOF, Albert-Einstein-Straße 7, 07745 Jena, Germany\\
\authormark{2}Abbe Center of Photonics, Friedrich Schiller University Jena, Albert-Einstein-Straße 6, 07745 Jena, Germany\\
\authormark{3}Cluster of Excellence Balance of the Microverse, Friedrich Schiller University Jena, Fürstengraben 1, 07743 Jena, Germany\\
\authormark{4}Institute for Applied Physics, Technical University of Darmstadt, Otto-Berndt-Straße 3, 64287 Darmstadt, Germany}

\email{\authormark{*}patrick.hendra@iof.fraunhofer.de} 


\begin{abstract*} 
We investigate the dual generation of type-0 and type-II spontaneous parametric down conversions (SPDCs) within a single periodically poled rubidium-doped KTiOPO$_4$ (PPRKTP) waveguide. By coupling a 45{\degree} linearly polarized pump laser into the waveguide, both SPDC processes are concurrently excited: the type-0 SPDC process is facilitated via third-order quasi-phase matching (QPM) utilizing the nonlinear coefficient $d_{33}$ , while the type-II SPDC process employs first-order QPM with the nonlinear coefficient $d_{24}$. This dual-SPDC scheme holds potential for applications in quantum communication protocols targeting the telecommunication wavelength.

\end{abstract*}

\section{Introduction}
In recent years, there has been growing interest in designing non-classical light sources for applications in quantum technology, including quantum sensing \cite{https://doi.org/10.1002/lpor.201900097} and quantum communication \cite{Chin:25}. While the use of bulk correlated sources is approaching optimal design and implementation, waveguide-based correlated sources offer advantages such as enhanced pair-flux by increasing the mode overlap between the pump, signal, and idler \cite{Fiorentino:07}. These sources can also be designed to be spatially and spectrally single-mode, enabling high-efficiency fiber coupling, while simultaneously achieving high spectral purity and brightness \cite{Steinlechner:14}---key parameters for high-rate, high-performance multi-photon experiments \cite{Meyer-Scott:18}. 

Multiple nonlinear optical parametric processes in the same bulk crystal or waveguide structure has long been recognized in the nonlinear optics community, particularly in periodically poled KTiOPO$_{4}$ (PPKTP) and periodically poled lithium niobate (PPLN) platforms \cite{Johnston:06, Briggs:21, Pysher:10, Lee:12, Chen:09}. These materials are favored due to their ferroelectric properties, which enable the formation of periodically poled domains for quasi-phase matching (QPM). By engineering specific poling periods, such crystals can simultaneously support multiple nonlinear interactions, including second harmonic generation (SHG) and spontaneous parametric down-conversion (SPDC), at desired wavelengths for a given pump wavelength and operating temperature. Prominent demonstrations include simultaneous type-0 and type-I SHG and their interference in bulk PPLN \cite{Johnston:06}, as well as type-I and type-II SHG in lithium niobate waveguides \cite{Briggs:21}. Concurrent type-0, type-I, and type-II SHG has also been realized in periodically poled KTP crystals \cite{Pysher:10}, while type-0 and type-II SPDC have been observed in both bulk \cite{Lee:12} and integrated undoped periodically poled KTP \cite{Chen:09}.

A recent advance in waveguide fabrication via ion exchange has enabled the use of rubidium-doped KTiOPO$_4$ (RKTP) as an alternative substrate \cite{Mutter:20}. This material exhibits low ionic conductivity, which facilitates the formation of homogeneous ferroelectric domains through electric field poling. The reduced ionic conductivity also permits the fabrication of quasi-phase-matched structures with submicron poling periods \cite{Mutter:22}, enabling the engineering of diverse biphoton states, including cascaded counter-propagating interactions \cite{Zukauskas:17} and backward-wave spontaneous parametric down-conversion (BW SPDC) \cite{Kuo:23}.


In this study, we investigate the spectral evolution and photon-pair flux of type-0 and type-II SPDC generated in a single periodically poled RKTP (PPRKTP) waveguide. We observe distinct spectral characteristics for the two SPDC types under varying crystal temperatures. Based on the experimental data, we determine the QPM orders corresponding to the type-0 and type-II SPDC processes.

\section{Experimental Setup}

\begin{figure}[htbp]
\centering\includegraphics[width=13.3cm]{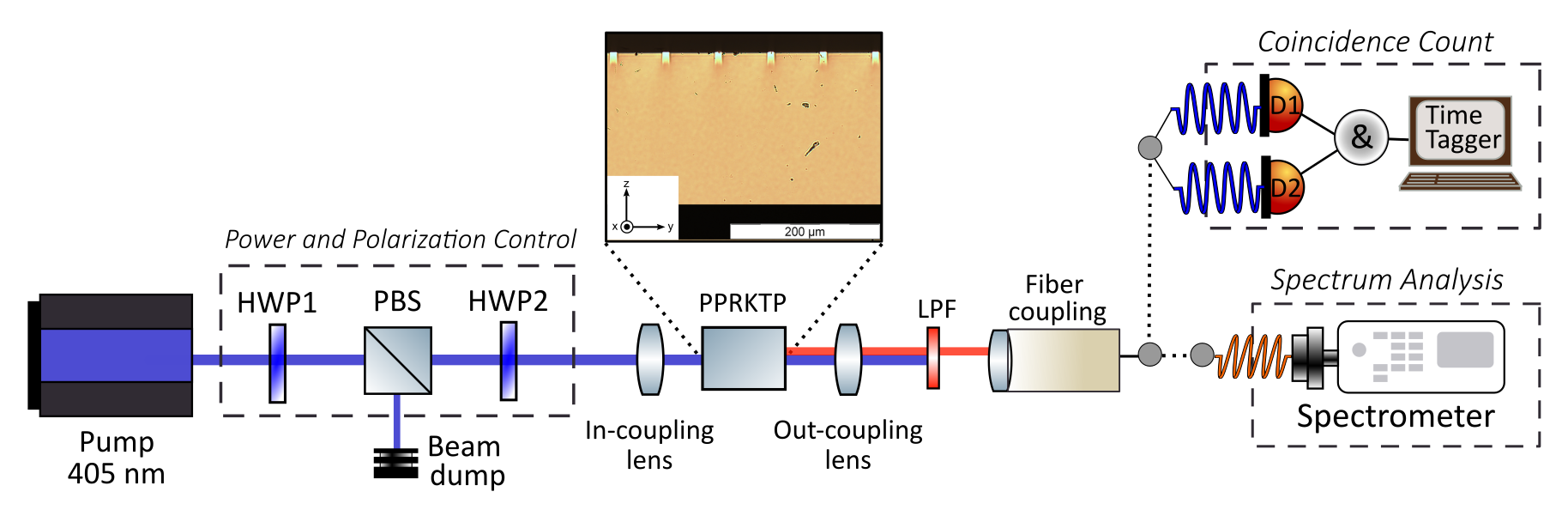}
\caption{Schematic of the experimental setup. A continuous-wave (CW) laser pumps the PPRKTP waveguide to generate both type-0 and type-II SPDC processes. The signal and idler photons are collected into a fiber, where their spectra and coincidence rates are analyzed.
}
\label{fig:2.1_ExperimentalSetup}
\end{figure}

The characterization setup is illustrated in the schematic diagram in Fig. \ref{fig:2.1_ExperimentalSetup}. A continuous-wave (CW) pump laser centered at 405 nm passes through a series of optical components for power and polarization control, including half-wave plates (HWP1 and HWP2) and a polarizing beam splitter (PBS). After polarization adjustment, the light is directed to an in-coupling lens that focuses the beam into the PPRKTP waveguide, where type-0 SPDC occurs for vertically polarized pump light and type-II SPDC occurs for horizontally polarized pump light.

A cross section of the PPRKTP input facet is shown in the inset of Fig. \ref{fig:2.1_ExperimentalSetup}, captured at 20X microscope magnification. The crystal axis is indicated in the bottom-left corner of the inset. The crystal dimensions are $\mathrm{x = 12\ mm}$, $\mathrm{y = 5.1\ mm}$, and $\mathrm{z = 1.0\ mm}$. The smallest channel width ($3\ \mathrm{\micro m}$) is located at the leftmost edge in the y-direction and increases in steps of $0.1\ \mathrm{\micro m}$ up to $7\ \mathrm{\micro m}$, resulting in a total of 41 channels. The periodically poled domains have a period of $9.96\ \mathrm{\micro m}$.

The output beam from the PPRKTP crystal is collected by an out-coupling lens and coupled into a fiber for subsequent analysis. Two low-pass filters (LPFs), each with a cut-off wavelength centered at $647\ \mathrm{nm}$, are used to suppress the transmitted pump beam. The spectral evolution of the generated photon pairs is characterized at different waveguide temperatures. Coincidence counting is also performed using a time tagger to estimate the photon-pair flux generated via SPDC in the PPRKTP waveguide.

For the purpose of characterization, we selected two arbitrary channels from the 41 available waveguide channels on the chip: channel number 6, with a width of $3.5\ \mathrm{\micro m}$, used for coincidence measurements, and channel number 11, with a width of $4\ \mathrm{\micro m}$, used for spectral characterization.


\section{Methods and Discussion}

\subsection{Temperature dependence of spectral evolution}

\begin{figure}[htbp]
\centering\includegraphics[width=13.3cm]{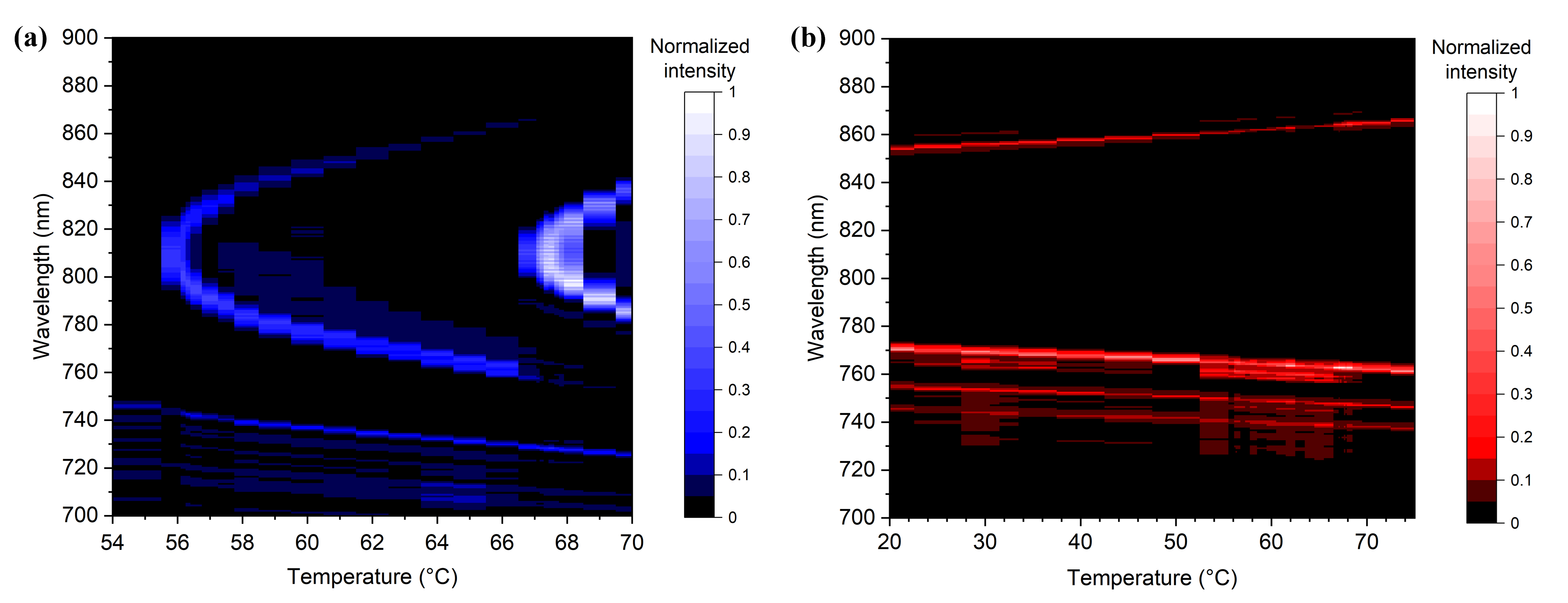}
\caption{Temperature-dependent spectral evolution of SPDC generated by waveguide channel number 11 (width: 4~$\micro$m). (a) Type-0 SPDC spectrum over a temperature range of $54{-}70\degree \mathrm{C}$. The brightest degenerate SPDC occurs near $67.5\degree \mathrm{C}$, with a secondary degenerate SPDC peak observed at $56\degree \mathrm{C}$. (b) Type-II SPDC spectrum measured over a temperature range of $20{-}75\degree \mathrm{C}$.}
\label{fig:3.1.1_WG11_SpecEvolution_HeatMap}
\end{figure}

Figure \ref{fig:3.1.1_WG11_SpecEvolution_HeatMap} illustrates the heatmap of the temperature-dependent spectral evolution of our SPDC source, comparing type-0 and type-II SPDC generated by waveguide number 11. Figure \ref{fig:3.1.1_WG11_SpecEvolution_HeatMap}(a) shows the spectral characteristics of type-0 SPDC, with intensity plotted as a function of wavelength and temperature. This data is obtained by pumping the waveguide with vertically polarized light. A prominent degenerate peak appears at approximately $67.5\degree \mathrm{C}$, indicating optimal phase matching at this temperature. Additionally, a secondary degenerate peak with lower brightness is observed around $56\degree \mathrm{C}$. The reduced intensity of this secondary peak suggests the involvement of a higher-order pump mode, which carries less energy than the fundamental mode and propagates within the waveguide, thereby generating lower SPDC photon flux \cite{Solntsev:18}. As the temperature increases, this secondary SPDC process undergoes a transition from degenerate to non-degenerate and eventually merges with the satellite peaks surrounding the main degenerate SPDC peak at $67.5\degree \mathrm{C}$. This behavior is consistent with prior studies on the spatial–spectral structure of waveguided SPDC, such as that reported by Mosley et al. \cite{Mosley:09}.

Figure \ref{fig:3.1.1_WG11_SpecEvolution_HeatMap}(b) presents the type-II SPDC spectrum, obtained by pumping the waveguide with horizontally polarized light. Unlike type-0 SPDC, the type-II process remains non-degenerate across the full temperature range. Here, the signal wavelength ($\lambda_\mathrm{s}$) remains below the degeneracy wavelength of 810~nm, ranging from 761~nm to 770~nm, while the idler wavelength ($\lambda_\mathrm{i}$) lies above 810~nm, ranging from 854~nm to 866~nm. The spectrum also exhibits a more stable non-degenerate profile, with minimal spectral drift across the temperature range.

Notably, the satellite peaks corresponding to non-degenerate idler wavelengths above 810~nm appear only faintly in both figures. This reduced visibility is attributed to the decreasing detection efficiency of the silicon-based spectrometer at infrared wavelengths, which limits the clarity of these spectral features.

\subsection{Photon-pair coincidence measurement}

\begin{figure}[htbp]
        \centering
        \includegraphics[width=13.3cm]{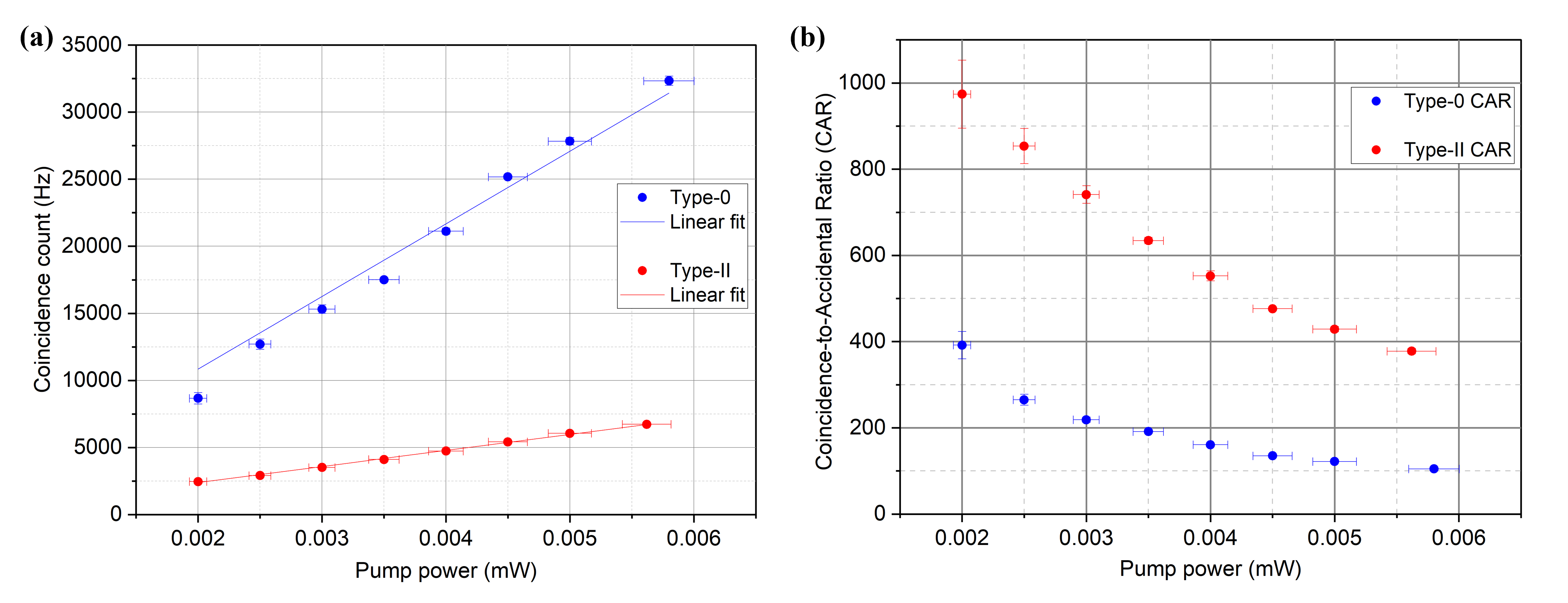}
        \caption{(a) Measured coincidence count rates for type-0 (blue dots) and type-II (red dots) SPDC as a function of input pump power, plotted together with their respective linear fits (blue and red lines). (b) Corresponding coincidence-to-accidental ratio (CAR) values as a function of pump power for type-0 (blue dots) and type-II (red dots) SPDC.}
        \label{fig:3.2.1_Coinc_CAR}
\end{figure}

The photon-pair generation rate is measured by splitting the SPDC beam using a 50:50 fiber beam splitter and directing each output arm to a single-photon detector. The arrival-time difference between photons in each arm is recorded using a time tagger with correlation electronics. The crystal temperature is set to \(63.5^{\circ} \mathrm{C}\), where the type-0 process exhibits degenerate signal and idler photons with \(\lambda_\mathrm{s} = \lambda_\mathrm{i} = 810\ \mathrm{nm}\). In contrast, type-II SPDC produces non-degenerate photon pairs with \(\lambda_\mathrm{s} = 762\ \mathrm{nm}\) and \(\lambda_\mathrm{i} = 865\ \mathrm{nm}\).

Figure~\ref{fig:3.2.1_Coinc_CAR} shows the photon-pair coincidence measurement for the PPRKTP waveguide. Coincidence counts are extracted from a coincidence histogram, which represents the distribution of time delays between photon detection events in two separate single photon detectors. Figure~\ref{fig:3.2.1_Coinc_CAR}(a) shows the dependence of the coincidence count on pump power, measured within a \(2\ \mathrm{ns}\) coincidence window corresponding to the full width at the base of this histogram. The coincidence count represents \emph{true coincidences}, calculated by subtracting accidental counts from the measured coincidences. A linear model is used to fit the measurement data, with the fit constrained to pass through the origin. The resulting fits yield high coefficients of determination: \(R^2 = 0.998\) for type-0 SPDC and \(R^2 = 0.999\) for type-II SPDC, indicating strong linearity in both cases. The slopes of the linear fits correspond to the photon-pair generation rate per milliwatt of pump power. Type-0 SPDC achieves a pair generation rate of \(5.417 \pm 0.097\ \mathrm{MHz\ mW^{-1}}\), while type-II SPDC reaches \(1.195 \pm 0.005\ \mathrm{MHz\ mW^{-1}}\), highlighting the significantly higher flux of type-0 SPDC.

Since a 50:50 beam splitter randomly routes both photons to the same or different detectors, only 50\% of pairs contribute to coincidences. Correcting for this, the \emph{effective} photon-pair rates are \(10.834 \pm 0.194\ \mathrm{MHz\ mW^{-1}}\) for type-0 and \(2.390 \pm 0.010\ \mathrm{MHz\ mW^{-1}}\) for type-II SPDC. Additionally, taking into account the losses in our system, such as pump coupling efficiency of 35\%, a single-photon detection efficiency of 65\% per detector arm, a single-mode fiber coupling efficiency of 30\%, and two long-pass filters with a transmission of 98\% each, the estimated intrinsic photon-pair generation rates are \(254.28\ \mathrm{MHz\ mW^{-1}}\) for type-0 and \(56.09\ \mathrm{MHz\ mW^{-1}}\) for type-II SPDC.

Among all the waveguide channels on the same chip, the highest observed pair rate was achieved using a waveguide with a width of \(4.0\ \mathrm{\mu m}\), yielding a type-0 SPDC rate of \(262.88\ \mathrm{MHz\ mW^{-1}}\). To enable fair comparison with prior works, we convert this result to a spectral density over a 20 nm filter bandwidth (equivalent to 9.14 THz). This yields an effective photon-pair rate of \(13.14\ \mathrm{MHz\ mW^{-1}\ nm^{-1}}\) or \(28.76 \times 10^{6}\ \mathrm{(s\cdot mW \cdot THz)^{-1}}\).

Table~\ref{tab:1_comparison} compares the photon-pair generation rates reported here with selected results from the literature, including sources based on PPLN, PPKTP, and AlGaAs. Compared to the dual-type PPKTP waveguides, our type-0 SPDC rate is approximately 2.86 times lower~\cite{Chen:09}. However, this difference is largely attributed to the longer waveguide length (15 mm) used in the cited work, making a direct comparison difficult.

Figure~\ref{fig:3.2.1_Coinc_CAR}(b) shows the coincidence-to-accidental ratio (CAR) as a function of pump power, where CAR is defined as the ratio of true coincidences to accidental counts. The data show an inverse relationship between CAR and pump power, indicating that increasing pump power raises the likelihood of multiphoton events, which in turn increases the number of accidentals. This trend is consistent with expectations, as higher pump powers lead to more background photons and multiphoton generation, reducing the signal-to-noise quality as reflected by the CAR.


\begin{table}[htbp]
\centering
\resizebox{\columnwidth}{!}{%
\begin{tabular}{|c|c|c|}
\hline
\rowcolor[HTML]{C0C0C0} 
\textbf{Reference} & \textbf{Platform}         & \textbf{Pair Generation Rate} \\ \hline
Fiorentino et al. (2007) \cite{Fiorentino:07}   & PPKTP bulk crystal        & 1.8 $\mathrm{MHz\ mW^{-1}}$                     \\ \hline
Fiorentino et al. (2007) \cite{Fiorentino:07}    & PPKTP waveguide           & 73 $\mathrm{MHz\ mW^{-1}\ nm^{-1}}$                       \\ \hline
Chen et al. (2009) \cite{Chen:09}       & Dual type PPKTP waveguide & \begin{tabular}[c]{@{}c@{}}Type-0 = 83 $\mathrm{MHz\ mW^{-1}\ THz^{-1}}$\\ Type-II = 250 $\mathrm{MHz\ mW^{-1}\ THz^{-1}}$\end{tabular} \\ \hline
Chen et al. (2019) \cite{Chen:19}       & PPLN nanowaveguide        &  69 $\mathrm{MHz\ mW^{-1}\ nm^{-1}}$                        \\ \hline
Zhao et al. (2020) \cite{Zhao:20}       & TF-PPLN waveguide         & 56 $\mathrm{MHz\ mW^{-1}\ nm^{-1}}$                     \\ \hline
Steiner et al. (2021) \cite{Steiner:21}    & AlGaAs-on-insulator       & 20 GHz                        \\ \hline
Zhang et al. (2023) \cite{Zhang:23}     & LNOI micro waveguide      & 180 $\mathrm{MHz\ mW^{-1}\ nm^{-1}}$                    \\ \hline
\textbf{This work} & \textbf{PPRKTP waveguide} & \begin{tabular}[c]{@{}c@{}} Type-0 = 13.14 $\mathrm{MHz\ mW^{-1}\ nm^{-1}}$\\  = 28.76 $\mathrm{MHz\ mW^{-1}\ THz^{-1}}$\end{tabular}           \\ \hline
\end{tabular}%
}
\caption{Comparison of photon-pair brightness from various reported sources in the literature with the results obtained in this work.}
\label{tab:1_comparison}
\end{table}

%


\subsection{Determining the QPM order}

The PPRKTP waveguide supports both type-0 and type-II phase matching, which can be switched by pumping the waveguide with vertically or horizontally polarized light, respectively. To further investigate the conditions under which phase matching is satisfied, both phase-matching equations for type-0 and type-II processes are solved simultaneously. The purpose of this analysis is to numerically determine unknown parameters in the phase-matching equations, such as the integer order of the grating harmonic and the waveguide dispersion contributions. The phase-matching equations for type-0 and type-II SPDC in the waveguide are given by:

\begin{align}
\text{Type-0} & \ \rightarrow \ \frac{2 \pi n_\mathrm{z} (\lambda_\mathrm{p}, T)}{\lambda_\mathrm{p}} - \frac{2 \pi n_\mathrm{z} (\lambda_\mathrm{s}, T)}{\lambda_\mathrm{s}} - \frac{2 \pi n_\mathrm{z} (\lambda_\mathrm{i}, T)}{\lambda_\mathrm{i}} - \frac{2 \pi m_\mathrm{x}}{\Lambda(T)} - k_{\text{wg}} = 0 \label{Eqn:PhaseMatching_Type0}\\
\text{Type-II} & \ \rightarrow \ \frac{2 \pi n_\mathrm{y} (\lambda_\mathrm{p}, T)}{\lambda_\mathrm{p}} - \frac{2 \pi n_\mathrm{z} (\lambda_\mathrm{s}, T)}{\lambda_\mathrm{s}} - \frac{2 \pi n_\mathrm{y} (\lambda_\mathrm{i}, T)}{\lambda_\mathrm{i}} - \frac{2 \pi m_\mathrm{y}}{\Lambda(T)} - k_{\text{wg}} = 0 \label{Eqn:PhaseMatching_TypeII}
\end{align}

where \( n_\mathrm{y} (n_\mathrm{z}) \) denotes the refractive index for y(z)-polarized light, \( m_\mathrm{x} (m_\mathrm{y}) \) is the grating harmonic order contributing to phase matching in type-0 (type-II) SPDC, \( \Lambda(T) \) is the temperature-dependent poling period, and \( k_{\mathrm{wg}} \) is the waveguide contribution to phase matching. The term \( k_{\mathrm{wg}} \) accounts for waveguide dispersion---i.e., the wavelength dependence of the propagation constant---arising from the specific geometry and material composition of the waveguide. These contributions affect the phase velocity of the interacting waves and thus influence the overall phase mismatch \cite{Ferreira:05}.

The wavelength and temperature dependent refractive index $n_{\mathrm{y,z}}(\lambda,T)$ is defined as:
\begin{equation}
    n_{\mathrm{y,z}}(\lambda,T) = n_{\mathrm{y,z}}(\lambda) + \Delta n(T,\lambda)
\end{equation}

where the refractive index $n_{\mathrm{y,z}}$ is given by the Sellmeier equation for KTP \cite{Fan:87}:
\begin{align}
    n_\mathrm{y}(\lambda) &= \sqrt{2.19229 + \frac{0.83547}{1 - 0.04970\ \lambda^{-2}} - 0.01621\ \lambda^2} \\
    n_\mathrm{z}(\lambda) &= \sqrt{2.25411 + \frac{1.06543}{1 - 0.05486\ \lambda^{-2}} - 0.02140\ \lambda^2}
\end{align}

And the temperature correction is:
\begin{equation}
    \Delta n(T, \lambda) = n_1(T - 25\,\degree\text{C}) + n_2(T - 25\,\degree\text{C})
\end{equation}

where $n_1$ and $n_2$ are given by:
\begin{equation}
    n_{1,2} = \sum_{m=0}^{3} \frac{a_m}{\lambda^m}
\end{equation}

The coefficients $a_m$ for y- and z-polarizations are found in Ref. \cite{Emanueli:03}.

As temperature increases, thermal expansion along the grating vector changes the poling period $\Lambda\left(T\right)$:
\begin{equation}
    \Lambda(T) = \Lambda_0 \left[1 + \alpha(T - 25\,\degree\text{C}) + \beta(T - 25\,\degree\text{C})^2 \right]
\end{equation}

where $\Lambda_0$ is the initial poling period at room temperature. The thermal expansion coefficients are $\alpha = 6.7 \times 10^{-6}\ \mathrm{\degree C}^{-1}$ and $\beta = 11 \times 10^{-9}\ \mathrm{\degree C}^{-2}$ \cite{Emanueli:03}.

\begin{figure}[htbp]
\centering\includegraphics[width=13cm]{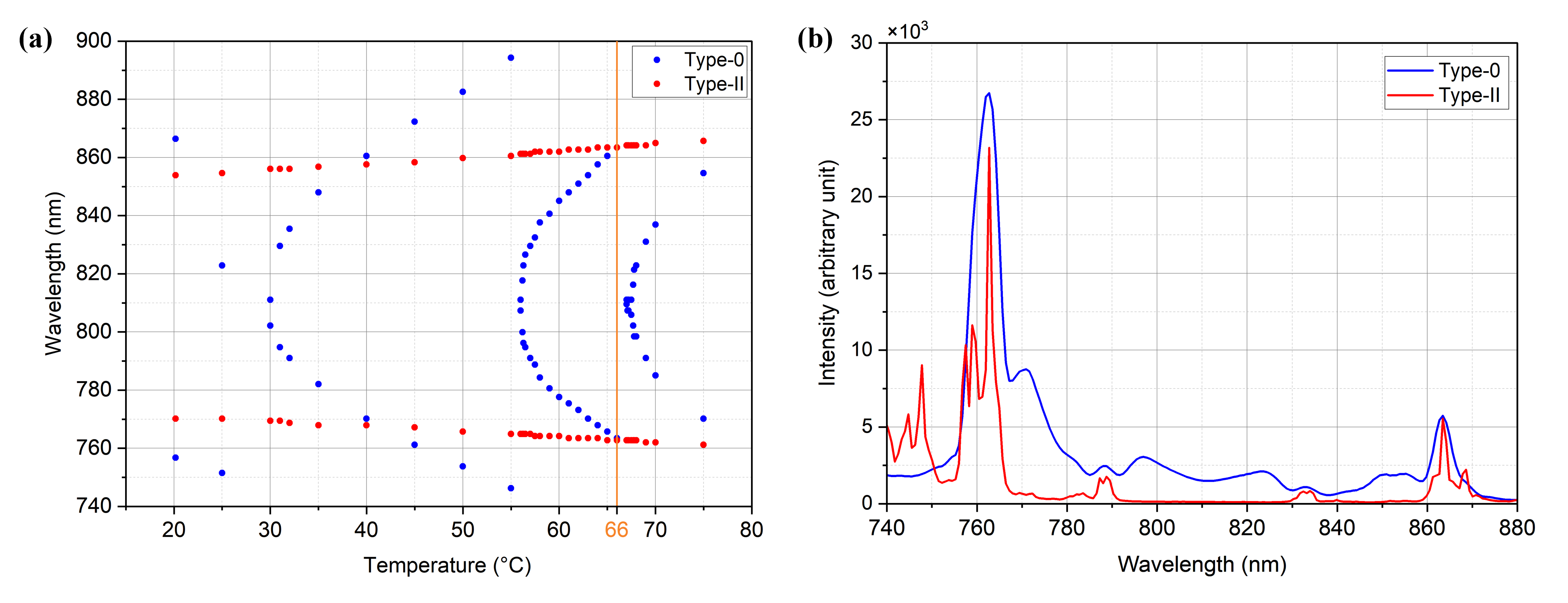}
\caption{(a) SPDC spectral overlap for type-0 (blue dots) and type-II (red dots). Each dot represents the peak intensity of the spectrum obtained from Fig.~\ref{fig:3.1.1_WG11_SpecEvolution_HeatMap}. The orange line indicates the intersection point of interest, occurring at a temperature of 66\degree C. (b) Spectral overlap between type-0 (blue curve) and type-II (red curve) at 66\degree C.}
\label{fig:3.3_Intersection}
\end{figure}

The wavelength values used in Eqs.~\ref{Eqn:PhaseMatching_Type0} and \ref{Eqn:PhaseMatching_TypeII} are extracted from overlapping experimental results presented in Fig.~\ref{fig:3.1.1_WG11_SpecEvolution_HeatMap}, as shown in Fig.~\ref{fig:3.3_Intersection}(a). From the figure, it can be observed that the signal and idler wavelengths for type-0 and type-II SPDC intersect at 66\degree C, corresponding to \(\lambda_{\text{s}} = 762.71~\mathrm{nm}\) and \(\lambda_{\text{i}} = 863.45~\mathrm{nm}\), as illustrated in Fig.~\ref{fig:3.3_Intersection}(b). This temperature is selected because it is the closest to the primary degenerate peak of type-0 SPDC, providing an optimal condition for comparison. We assume that \(k_{\mathrm{wg}}\) is the same in both Eq.~\ref{Eqn:PhaseMatching_Type0} and Eq.~\ref{Eqn:PhaseMatching_TypeII}, since the signal and idler wavelengths are identical for both SPDC types at this temperature.



Inserting the poling period $\Lambda = 9.96\,\mathrm{\mu m}$ and these wavelengths into Eqs.~\ref{Eqn:PhaseMatching_Type0} and \ref{Eqn:PhaseMatching_TypeII} yields:
\begin{align}
    \text{Type-0} &\ \rightarrow\ 30.47 = 28.64 + 0.63 \, m_\mathrm{x} + k_{\text{wg}} \label{Eqn:type-0} \\
    \text{Type-II} &\ \rightarrow\ 28.57 = 28.00 + 0.63 \, m_\mathrm{y} + k_{\text{wg}} \label{Eqn:type-II}
\end{align}

Subtracting Eq.~\ref{Eqn:type-II} from Eq.~\ref{Eqn:type-0} gives:
\begin{equation}
    m_\mathrm{y} = m_\mathrm{x} - 2.01
\end{equation}

Applying constraints:
\begin{itemize}
    \item $m_{\mathrm{x,y}} \in \mathbb{Z}^+$: the poling order must be a positive integer.
    \item $m_{\mathrm{x,y}} = 2k + 1$ for $k \in \mathbb{Z}$: only odd orders contribute to maximum diffraction efficiency \cite{Pandiyan:08, Mutter_thesis:22}.
\end{itemize}

The valid solution is $m_\mathrm{x} = 3$ and $m_\mathrm{y} = 1$, meaning type-0 uses third-order and type-II uses first-order poling---yielding a waveguide phase mismatch of $k_\mathrm{wg} = -0.056\mathrm{\mu m}^{-1}$.

While third-order QPM reduces conversion efficiency compared to first-order due to insufficient compensation of the down-converted field's oscillation \cite{Sokolovskaya:2016}, the higher nonlinear coefficient $d_{33} = 18.5\,\mathrm{pm/V}$ used in type-0 compensates for this. Type-II, using $d_{24} = 3.92\,\mathrm{pm/V}$ (measured at 880\,nm) \cite{Vanherzeele:92}, exhibits lower brightness. This explains the comparatively lower brightness in our work versus Chen et al. \cite{Chen:09}, where type-0 used zeroth-order and type-II used first-order QPM.


\section{Conclusion}
In conclusion, we have successfully demonstrated the simultaneous generation of dual type-0 and type-II spontaneous parametric down-conversion (SPDC) in a periodically poled, rubidium-doped KTP waveguide. By employing a 45° linearly polarized pump laser, third-order quasi-phase matching (QPM) is achieved for type-0 SPDC via the nonlinear coefficient $d_{33}$, and first-order QPM for type-II SPDC via $d_{24}$.

Spectral characterization reveals that the type-0 process exhibits multiple degeneracies, whereas the type-II process remains non-degenerate across the scanned temperature range of $20{-}75\degree \mathrm{C}$. The highest photon-pair flux was observed in the type-0 configuration, reaching an effective coincidence count rate of 13.14 $\mathrm{MHz\ mW^{-1}\ nm^{-1}}$, or equivalently, 28.76 $\mathrm{MHz\ mW^{-1}\ THz^{-1}}$.

The use of rubidium-doped KTP is particularly noteworthy, as it significantly reduces ionic conductivity. This enables lower poling periods and minimizes poling errors, both of which are crucial for achieving precise QPM.

These findings contribute to a deeper understanding of SPDC mechanisms in engineered nonlinear optical media and pave the way for compact, efficient photon-pair sources operating in the first telecom window—relevant for applications in quantum communication. Future work will aim to further optimize these SPDC processes and explore their integration into scalable quantum photonic platforms.

\begin{backmatter}
\bmsection{Funding} This work was funded by the Deutsche Forschungsgemeinschaft (DFG, German Research Foundation) under Germany´s Excellence Strategy – EXC 2051 – Project-ID 390713860, by the European Union’s Horizon 2020 Research and Innovation Action under Grant Agreement No. 101113901 (Qu-Test, HORIZON-CL4-2022-QUANTUM-05-SGA) and by a grant funded by the Federal Ministry of Education and Research (QUANTIFISENS - 03RU1U071M).
 
\bmsection{Acknowledgments} The authors gratefully acknowledge O. Boucher for assistance in the laboratory and C.~Canalias for supplying the waveguide. The authors thank S. Sharma and F. Steinlechner for facilitating the collaboration that made this research possible.
 
\bmsection{Disclosures} The authors declare no conflict of interest.
 
\bmsection{Data availability} The data that support the findings of this study are available from the corresponding author, P.H., upon reasonable request.

\end{backmatter}

\bibliography{Optica-template}






\end{document}